\pdfoutput=1
\documentclass[preprint]{elsarticle}
\makeatletter
\def\ps@pprintTitle{
	\let\@oddhead\@empty
	\let\@evenhead\@empty
	\def\@oddfoot{\centerline{\thepage}}
	\let\@evenfoot\@oddfoot}
\makeatother

\usepackage{amsmath}
\usepackage{amssymb}
\usepackage{xcolor}
\usepackage{bbm}
\usepackage{epstopdf}
\usepackage{epsfig}
\usepackage{url}
\usepackage{stmaryrd}
\usepackage{siunitx}
\usepackage{hyperref}
\usepackage{mathtools}

\begin{document}
\begin{frontmatter}
\title{The \textit{transcoding sampler} for stick-breaking inferences on Dirichlet process mixtures\tnoteref{t1}}
\tnotetext[t1]{This article is based on chapter 7 of my PhD thesis at Durham University \cite{myOwn}.}

\author{Carlo Vicentini}
\ead{carlo.vicentini@mathstat.net}

\begin{abstract}
Dirichlet process mixture models suffer from slow mixing of the MCMC posterior chain produced by stick-breaking Gibbs samplers, as opposed to collapsed Gibbs samplers based on the Polya urn representation which have shorter integrated autocorrelation time (IAT).
	
We study how cluster membership information is encoded under the two aforementioned samplers, and we introduce the \textit{transcoding algorithm} to switch between encodings. We also develop the \textit{transcoding sampler}, which consists of undertaking posterior partition inference with any high-efficiency sampler, such as collapsed Gibbs, and to subsequently transcode it to the stick-breaking representation via the transcoding algorithm, thereby allowing inference on all stick-breaking parameters of interest while retaining the shorter IAT of the high-efficiency sampler. 
	
The transcoding sampler is substantially simpler to implement than the slice sampler, it can inherit the shorter IAT of collapsed Gibbs samplers and it can also achieve zero IAT when paired with a posterior partition sampler that is i.i.d., such as the sequential importance sampler.
\end{abstract}

	
\end{frontmatter}

\section{Introduction}\label{sec1}
Posterior inference of the Dirichlet process mixture model parameters under the stick-breaking representation is known to suffer from slowdowns caused by difficulties in the Gibbs sampler moving between local modes. This class of samplers is known as \textit{conditional} samplers, or \textit{stick-breaking} samplers. Conversely, \textit{marginal} Gibbs samplers (also called \textit{collapsed} samplers) based on the Polya urn representation of the Dirichlet process operate in a smaller parameter space, as they marginalise the Dirichlet process out; they are known to be less affected by the issue, and to have shorter integrated autocorrelation time (IAT). They also use a different integer encoding for cluster membership indicators, as opposed to conditional samplers. Various Metropolis jumps have been proposed to induce the stick-breaking sampler to move more frequently between local modes, thereby attaining faster mixing; however, none of these methodologies has so far reached the same level of performance as the marginal samplers. Key references in this respect are \cite{retrospective} and \cite{hastie_sampling_2015}.

We observe a gap in literature pertaining to the study of the relationship between the collapsed parameter space of the Polya urn, and the wider parameter space of the stick-breaking representation. The former leads to cluster labels which are numbered in order of appearance, and which carry no specific meaning as they are exchangeable; the latter instead carry very specific meaning, as they point to which stick in the stick-breaking construction each observation originates from. We discuss the relationship between the two encodings, and we devise a way to infer one from the other, in both directions, to ultimately design a new sampler which can augment the partition posterior obtained under either encoding, with inferences pertaining to all stick-breaking parameters of interest.

This article is structured as follows. Section \ref{sec:encodings} discusses and contrasts the two encoding conventions. Section \ref{sec:transcoding} discusses how to transcode between encodings, either deterministically, or through statistical inference via the \textit{transcoding algorithm}.  Section \ref{sec:tranSampler} builds on the \textit{transcoding algorithm} and introduces the \textit{transcoding sampler}, for full posterior inference of all stick-breaking parameters; it also discusses its performance. Section \ref{sec:compare} offers a summary of the relationship between this article and other work in this area, while section \ref{sec:conclusions} draws the conclusions.

\section{Encodings}\label{sec:encodings}
In statistics, information about data and parameters pertaining to a model must be encoded into standard format according to some convention, for the information to be useable. While doing so is often straightforward, the numerous alternate representations of the Dirichlet process (henceforth DP) offer a multiplicity of options to choose from, the relationships between which are complex, especially when it pertains to identifying partitions. This is partly inherited from the fact that partitions themselves, as mathematical objects, feature several competing coding conventions, and it is also partly due to the exchangeability properties often enjoyed by the many alternate representations of the DP.

In this section we discuss two encoding methods for DP cluster membership indicators, and how they are equivalent in terms of the partition that they identify, yet one holds more informative power than the other with respect to the information that it implicitly carries.

Consider a model of $n$ observations from a Dirichlet Process Mixture (henceforth DPM), written as follows: 
\begin{align}\label{eq:DPMb}
	&y_i\mid \theta_i \sim p\left(y_i\mid\theta_i\right),\\
	&\theta_i\mid G \sim G,\nonumber\\
	&G\sim \textrm{DP}\left(\alpha,G_0\right),\nonumber
\end{align}
where $\boldsymbol{y}=\left(y_1,\ldots,y_n\right)$ is a vector of $n$ observations while $\boldsymbol{\theta}=\left(\theta_1,\ldots,\theta_n\right)$ is the vector of their latent parameters; $G$ indicates both the Dirichlet process and its realisation, $G_0$ is called the base measure, and $\alpha$ is the precision parameter. 

Repeated values of $\theta_i$ identify random clusters of observations; often, inference on these clusters is one of the primary objectives of resorting to a DPM model in the first place. It is common practice to augment the model specification from equation \ref{eq:DPMb} with additional latent variables, to express cluster membership information in isolation from the locations of the atoms, for example to enable algorithms which are primarily based on the random partition induced by the DP, or to accompany applications where the location of the atoms is not of primary interest. There are multiple ways and conventions to encode cluster membership information into the latent cluster membership indicator vector.

\subsection{Encoding in order of appearance}\label{sec:polya-encoding}
The original article pertaining to the Polya urn representation of the Dirichlet process \citep{blackwellMacQueen} ignores the topic of the location of the atoms of the DP, and only focuses on its probability masses and on the partition that the Polya urn sequence induces. The article only refers to the balls in the Polya urn having a certain \textit{colour} $x$, and it does not point to any specific labelling convention for it. In this respect, $x$ could belong to any set, as long as the set is countably infinite; in principle the set could even be a set of words or descriptors, to identify infinite colours, although doing so would be impractical, as no natural language that we know of has an infinite set of descriptors to match an infinite palette.

A more practical way, although certainly not the only way, to encode the information about the colours of the balls extracted from the urn is to rely on natural numbers. This is for example how \cite{de_finetti_theory_2017}
discusses the $2$-colour Polya urn, where he assigns $1$ to indicate a white ball, and $0$ to indicate black. Similarly, colour membership in an $n$-colour Polya urn can be expressed with integers from $\mathbb{N}_n$, to map to specific colours. However, as outlined, the strategy of listing all possible colours and mapping them to $\mathbb{N}$ does not work well with the Dirichlet process due to its infinite dimensionality.

Common practice in the infinite-dimensional Polya urn setting is therefore to abandon the colour analogy entirely, and to encode cluster membership via $\mathbb{N}$, with the clusters labelled in the order whereby they appear in the sampling process. Given an $n$-dimensional sample, cluster membership is encoded via vector $\boldsymbol{s}=\left(s_1,\ldots,s_n\right)$, defined so that $A_j\equiv \left\{i:s_i=j, i\in \mathbb{N}_n\right\}$. This results in a scheme where $s_1=1$, always; $s_2=1$ if the second ball from the urn is the same colour as the first ball, and $s_2=2$ otherwise, and so on:
\begin{align}
	&p\left(s_1=1\right)=1,\nonumber\\
	&p\left(s_i\mid s_{i-1},\ldots,s_1\right)=\frac{\alpha}{\alpha+i-1}\delta_{k_{i-1}+1}\left(s_i\right)+\sum_{l=1}^{k_{i-1}} \frac{n_{i-1,l}}{\alpha+i-1}\delta_l\left(s_i\right),\label{eq:polya10}
\end{align}
where $k_{i-1}$ and $n_{i-1,l}$ are respectively the number of clusters and the size of cluster $l$ over the first $i-1$ observations.

\subsection{Encoding in stick-breaking order}
The phrase \textit{stick-breaking} refers to the following construction, where $w_1,w_2,\ldots$ can be seen as pieces broken out of a stick of length $1$:
\begin{align}\label{eq:sethuraman}
	&w_1:=v_1,\nonumber\\
	&w_h:=v_h\prod_{l<h}\left(1-v_l\right),\quad h=2,3,\ldots\nonumber\\
	&v_h\sim \textrm{Beta}\left(1,\alpha\right), \quad h=1,2,\ldots,\\
	&m_h \sim G_0.\nonumber
\end{align}
It can be proved \citep{sethuraman} that the distribution of the following random measure is $\textrm{DP}\left(\alpha,G_0\right)$:
\begin{equation}\label{eq:stick}
	G:=\sum_{h=1}^{\infty}w_h \delta_{m_h}.
\end{equation}

In the DP literature, we mainly see two approaches to encoding the clustering scheme arising from the stick-breaking process. The first assigns integers to the cluster membership indicator vector from $\mathbb{N}_k$ in the order of appearance of the sticks that the observations are sampled from. For example, conditional to $\boldsymbol{w}$, if the first drawn observation  originates from the stick of length $w_h$, then it is assigned the label $1$, and so is any other subsequent draw from the same stick; the next observation to be drawn from any stick other than $w_h$, and any other subsequent observations drawn from the same, are assigned the label $2$, and so on. The resulting sequence of cluster membership indicators is clearly equivalent to the one arising from the Polya urn, and it is labelled \textit{in order of appearance} (the same discussed in section \ref{sec:polya-encoding}). For example, this type of encoding is used in the many influential articles from Pitman, where he discusses the stick-breaking process (for example, see Section 3.1 of \cite{pitman_combinatorial_2002}).

The other approach involves indexing the sticks in their order of construction as per equation \ref{eq:sethuraman}, and then assigning to each observation the index of the stick that the observation was sampled from. In symbols, consider equation \ref{eq:sethuraman}, and express cluster membership through the vector $\boldsymbol{r}=\left(r_1,\ldots,r_n\right)$, where
\begin{equation*}
	p\left(r_i=h\mid\boldsymbol{w}\right) =w_h, \quad h=1,2,\ldots
\end{equation*}
While it is ubiquitous in the Bayesian nonparametric literature, this encoding method carries no name to identify it, and we hereby name it \textit{stick-breaking order}. 

We observe that encoding in stick-breaking order carries more information than encoding in order of appearance, as the latter is anyway already implicitly included in the presentation order of the stick-breaking indices\footnote{This is on the assumption that the order of appearance of each data point is known, and reflected in the data set.}. While an ordered vector $\boldsymbol{r}$ encoded in stick-breaking order can be deterministically brought back to its order of appearance ($\boldsymbol{s}$), by checking which elements of $\boldsymbol{r}$ are new as one progresses through the ordered sequence, the reverse is not true, and inference is needed to attempt to derive $\boldsymbol{r}$ back from $\boldsymbol{s}$.

\subsection{Size-biased random permutations}\label{section:sizeBiased}
The concept of \textit{order of appearance} is strictly connected to that of size-biasing. A size-biased random permutation $\tilde{\boldsymbol{w}}=\left(\tilde{w}_1,\tilde{w}_2,\ldots\right)$ of $\boldsymbol{w}=\left(w_1,w_2,\ldots\right)$ is defined as follows (see for example \cite{phadia_prior_2015}). First sample $\tilde{w}_1$ from
\begin{equation*}
	p\left(\tilde{w}_1=w_h\mid \boldsymbol{w}\right)=w_h,	\quad h=1,2,\ldots,
\end{equation*}
and then sample $\tilde{w}_2,\tilde{w}_3,\ldots$ from
\begin{equation}\label{eq:sizeBias2}
	p\left(\tilde{w}_{k+1}=w_h\mid\tilde{w}_1,\ldots,\tilde{w}_{k},\boldsymbol{w}\right)=\frac{w_h \ \mathbbm{1}\left[w_h\neq \tilde{w_i}, \textrm{for } 1\leq i \leq k\right]}{1-\tilde{w}_1-\tilde{w}_2-\ldots-\tilde{w}_k},\ h=1,2,\ldots
\end{equation}
for $k \geq 1$. 

An equivalent definition \citep{pitman_random_1996} involves sampling the indices $I_1,I_2,\ldots$ from the categorical distribution with parameter $\boldsymbol{w}$, and denoting the distinct values in $I_1,I_2,\ldots$, in order of appearance, as $\tilde{I}_1,\tilde{I}_2,\ldots$. Then $\left(\tilde{w}_1,\tilde{w}_2,\ldots\right):=\left(w_{\tilde{I}_1},w_{\tilde{I}_2},\ldots\right)$ is a size-biased permutation of $\boldsymbol{w}$.

An important property of $\boldsymbol{w}$ is that it is invariant to size-biased random permutations \citep{arratia_logarithmic_2003, pitman_random_1996, engen_note_1975, mccloskey_model_1965}, meaning that the random variable $\boldsymbol{w}$ and its size-biased random permutation $\tilde{\boldsymbol{w}}$ are equal in distribution:
\begin{equation*}
	\boldsymbol{w}\,{\buildrel d \over =}\,\tilde{\boldsymbol{w}}.
\end{equation*}
Because it has the same probability distribution as that of its random size-biased permutation, the sequence $w_1,w_2,\ldots$ is said to be \textit{in size-biased order}. The random variable arising from any number of repeated size-biased random permutations of $\boldsymbol{w}$ still has the same probability distribution as $\boldsymbol{w}$.

\section{Transcoding}\label{sec:transcoding}
In this section we describe how to transcode cluster membership indicators from one encoding to the other. Order of appearance labels can deterministically be obtained from stick-breaking labels, while the reverse is less straightforward and requires statistical inference.

\subsection{From stick-breaking to order of appearance}\label{section:r2s-1}
Recall that repeated values in $\boldsymbol{r}$ induce a random partition $\left\{A_1,\ldots,A_k\right\}$ on $\mathbb{N}_n$, where we define $A_1,\ldots,A_k$ in order of appearance, such that $1 \in A_1$ and for each $2\leq i\leq k$ the first element of $\mathbb{N}_n\setminus \left(A_1 \cup \ldots \cup A_{i-1}\right)$ belongs to $A_i$. The vector $\boldsymbol{s}$ is then entirely determined by $\boldsymbol{r}$ through the mapping $g:\mathbb{N}^n \mapsto \mathbb{N}_k^n$ induced by the relationship:
\begin{equation*}
	s_i=j\iff i\in A_j.
\end{equation*}

\subsection{From order of appearance to stick-breaking -- the transcoding algorithm}\label{sec:transAlgo}
In a set of $n$ observations with $K=k_n$ observed clusters, denote the distinct values of $\boldsymbol{r}$ (in the order whereby they appear in the vector $\boldsymbol{r}$) by $\left(r^\star_1,\ldots,r^\star_{k_n}\right)$, as follows:
\begin{align*}
	r_1^\star&=r_1,\\
	r_j^\star&=r_{\min\left\{i \in \mathbb{N}_n:\ r_i \notin \left\{r_1^\star,\ldots,r_{j-1}^\star\right\}\right\}}, \quad j=2,\ldots,k_n.
\end{align*}
The random vector $\boldsymbol{r}^\star$ carries precisely the extra information that is needed over $\boldsymbol{s}$ to retrieve $\boldsymbol{r}$, as the pair $\left(\boldsymbol{r}^\star,\boldsymbol{s}\right)$ is informationally equivalent to $\boldsymbol{r}$. We further denote by $\tilde{\boldsymbol{w}}$ the lengths of the sticks in the order whereby the original sticks are discovered in the data sampling process. We also use the symbols $\tilde{\boldsymbol{n}}=\left(\tilde{n}_1,\ldots,\tilde{n}_k\right)$ to distinguish the cluster sizes in order of appearance from $\boldsymbol{n}=\left(n_1,n_2,\ldots\right)$, the cluster sizes indexed to match $\boldsymbol{w}$.

The simplest approach to infer $\boldsymbol{r}$ from $\boldsymbol{s}$ involves the accept/reject algorithm, which we present several increasingly efficient approaches for in \ref{appendix:AR}; however, even in its most efficient form, it still proves to be wasteful, to the point of being impractical with large data sets, which motivates the approach below (the \textit{transcoding algorithm}), which augments the target space to allow sampling with almost sure acceptance.

We introduce the indicator variable $\boldsymbol{t}=\left(t_1,t_2,\ldots\right)$ to explicitly describe the order of appearance of $\left(w_1,w_2,\ldots\right)$:
\begin{equation}\label{eq:t-r}
	t_j=i\Leftrightarrow r_i^\star=j.
\end{equation}
Our motivating idea is that the pair $\left(\boldsymbol{\tilde{w}},\boldsymbol{t}\right)$ expresses the same information as $\left(\boldsymbol{w},\boldsymbol{r}^\star\right)$; in doing so, we took inspiration from section 3.2 of \cite{lee_defining_2013}, and we reversed it.

Now that $\boldsymbol{t}$ is defined, we can draw $\boldsymbol{r}^\star\mid\boldsymbol{s}$ by sampling from the joint posterior $p\left(\boldsymbol{r}^\star,\boldsymbol{w},\boldsymbol{t},\tilde{\boldsymbol{w}}\mid\boldsymbol{s}\right)$:
\begin{align}
	p\left(\boldsymbol{r}^\star,\boldsymbol{w},\boldsymbol{t},\tilde{\boldsymbol{w}}\mid\boldsymbol{s}\right)&=p\left(\boldsymbol{r}^\star\mid\boldsymbol{w},\boldsymbol{t},\tilde{\boldsymbol{w}},\boldsymbol{s}\right)\cdot p\left(\boldsymbol{w},\boldsymbol{t},\tilde{\boldsymbol{w}}\mid\boldsymbol{s}\right)\nonumber\\
	&=p\left(\boldsymbol{r}^\star\mid\tilde{\boldsymbol{w}},\boldsymbol{t}\right)\cdot p\left(\boldsymbol{w},\boldsymbol{t}\mid\tilde{\boldsymbol{w}},\boldsymbol{s}\right)\cdot p\left(\tilde{\boldsymbol{w}}\mid\boldsymbol{s}\right)\nonumber\\
	&=p\left(\boldsymbol{r}^\star\mid\tilde{\boldsymbol{w}},\boldsymbol{t}\right)\cdot p\left(\boldsymbol{w},\boldsymbol{t}\mid\tilde{\boldsymbol{w}}\right)\cdot p\left(\tilde{\boldsymbol{w}}\mid\boldsymbol{s}\right),\label{eq:s2r}
\end{align}
where the first term in equation \ref{eq:s2r}  is an indicator function which does not require any sampling, as $\boldsymbol{r}^\star\mid \tilde{\boldsymbol{w}},\boldsymbol{t}$ is entirely determined by $\left(\tilde{\boldsymbol{w}},\boldsymbol{t}\right)$, while the second and third term need to be sampled.

The algorithm involves three steps (in the reverse order of the three factors from equation \ref{eq:s2r}):
\begin{enumerate}
	\item sample $\tilde{\boldsymbol{w}}\mid\boldsymbol{s}$,
	\item sample $\boldsymbol{w},\boldsymbol{t}\mid\tilde{\boldsymbol{w}}$,
	\item deterministically obtain $p\left(\boldsymbol{r}^\star\mid\tilde{\boldsymbol{w}},\boldsymbol{t}\right)$ and therefore $\boldsymbol{r}\mid\boldsymbol{w},\boldsymbol{t},\tilde{\boldsymbol{w}},\boldsymbol{s}$.
\end{enumerate}

Although the algorithm does entail infinite-length vectors, from a practical perspective only a finite number of elements needs to be drawn, hence the algorithm can still be executed exactly.

\subsubsection{Sampling \texorpdfstring{$\tilde{\boldsymbol{w}}\mid \boldsymbol{s}$}{w.tilde given s}}
The posterior distribution of $\tilde{\boldsymbol{w}}\mid\boldsymbol{s}$ can be expressed as
\begin{align}
	&p\left(\tilde{\boldsymbol{w}}\mid\boldsymbol{s}\right)=\frac{p\left(\boldsymbol{s}\mid\tilde{\boldsymbol{w}}\right)p\left(\tilde{\boldsymbol{w}}\right)}{p\left(\boldsymbol{s}\right)}\nonumber\\
	&\begin{multlined}=\frac{1}{p\left(\boldsymbol{s}\right)}\left(\prod_{i=1}^k \tilde{w}_i^{\tilde{n}_i-1}\right)\left(\prod_{i=1}^{k-1}\left(1-\sum_{j=1}^i \tilde{w}_j\right)\right)\\
		\frac{\alpha^k\left(1-\tilde{w}_1-\ldots-\tilde{w}_k\right)^{\alpha-1}}{\left(1-\tilde{w}_1\right)\ldots\left(1-\tilde{w}_1-\ldots-\tilde{w}_{k-1}\right)}\end{multlined}\nonumber\\
	&=\frac{\alpha^k}{p\left(\boldsymbol{s}\right)}\left(\prod_{i=1}^k \tilde{w}_i^{\tilde{n}_i-1}\right)\left(1-\tilde{w}_1-\ldots-\tilde{w}_k\right)^{\alpha-1},\label{eq:wtilde}
\end{align}
where the second passage above leverages corollary 7 from \cite{pitman_exchangeable_1995} and the fact that:
\begin{equation}\label{eq:pitman0}
	p\left(\tilde{n}_1,\ldots,\tilde{n}_k\mid \tilde{\boldsymbol{w}}\right)=\left(\prod_{i=1}^k \tilde{w}_i^{\tilde{n}_i-1}\right)\left(\prod_{i=1}^{k-1}\left(1-\sum_{j=1}^i \tilde{w}_j\right)\right).
\end{equation}

We recall that $\boldsymbol{w}$ and $\tilde{\boldsymbol{w}}$ are equal in distribution. As the former can be expressed in terms of $\boldsymbol{v}$, where each element is a priori beta-distributed, then the latter also can, and we name the equivalent terms $\tilde{\boldsymbol{v}}$.

Recalling that the determinant $\lvert J\rvert$ of the Jacobian of the transform
\begin{equation*}
	v_i=\frac{w_i}{1-\sum_{l<i}w_l}
\end{equation*}
is
\begin{equation*}
	\lvert J\rvert=\prod_{i=1}^{k-1}\left(1-v_i\right)^{k-i},
\end{equation*}
we apply it to derive $p\left(\tilde{\boldsymbol{v}}\mid\boldsymbol{s}\right)$ from $p\left(\tilde{\boldsymbol{w}}\mid\boldsymbol{s}\right)$:
\begin{align*}
	p\left(\tilde{\boldsymbol{v}}\mid\boldsymbol{s}\right)&=\frac{\alpha^k}{p\left(\boldsymbol{s}\right)}\left(\prod_{i=1}^k \tilde{v}_i^{\tilde{n}_i-1}\prod_{l<i}\left(1-\tilde{v}_l\right)^{\tilde{n}_i-1}\right) \left(\prod_{i=1}^k 1-\tilde{v}_i\right)^{\alpha-1} \prod_{i=1}^{k-1}\left(1-\tilde{v}_i\right)^{k-i}\\
	&=\frac{\alpha^k}{p\left(\boldsymbol{s}\right)} \left(\prod_{i=1}^k \tilde{v}_i^{\tilde{n}_i-1}\prod_{l<i}\left(1-\tilde{v}_l\right)^{\tilde{n}_i-1}\right)  \left(1-\tilde{v}_k\right)^{\alpha-1}\prod_{i=1}^{k-1}\left(1-\tilde{v}_i\right)^{\alpha+k-i-1}\\
	&=\frac{\alpha^k}{p\left(\boldsymbol{s}\right)} \left(\prod_{i=1}^k \tilde{v}_i^{\tilde{n}_i-1}\right)\prod_{i=1}^k \left(1-\tilde{v_i}\right)^{\sum_{l>i}\tilde{n}_l+\alpha-1},
\end{align*}
which leads to:
\begin{equation*}
	\tilde{v}_i\mid\boldsymbol{s}\sim \textrm{Beta}\left(\tilde{n}_i,\alpha+\sum_{l>i}\tilde{n}_l\right).
\end{equation*}
This provides the joint distribution of $\tilde{\boldsymbol{w}}\mid\boldsymbol{s}$ and $\tilde{\boldsymbol{v}}\mid\boldsymbol{s}$. Their marginals are  consistent (as they should be) with the result from \cite{hoppe_sampling_1987}, according to which $\tilde{w}_1\mid\boldsymbol{s}\sim \textrm{Beta}\left(\tilde{n}_1,\alpha+n-\tilde{n}_1\right)$ and, by exchangeability,
\begin{equation*}
	\tilde{w}_i\mid\boldsymbol{s}\sim \textrm{Beta}\left(\tilde{n}_i,\alpha+n-\tilde{n}_i\right),
\end{equation*} 
as can also be obtained in our setting from equation \ref{eq:wtilde} by integrating out the nuisance terms. This also dovetails with theorem 1 of \cite{ferguson1973} (also explained in theorem 1 and corollary 20 of \cite{pitman_developments_1996}, which better contextualises it with $\tilde{\boldsymbol{w}}$), according to which
\begin{equation}\label{eq:fall}
	\left(\tilde{w}_1,\ldots,\tilde{w}_k,1-\sum_{j=1}^k \tilde{w}_j\mid  \tilde{n_1},\ldots,\tilde{n_k}\right) \sim \textrm{Dirichlet}\left(\tilde{n}_1,\ldots,\tilde{n}_k,\alpha\right),
\end{equation}
by definition of a Dirichlet process. 

\subsubsection{Sampling \texorpdfstring{$\boldsymbol{w},\boldsymbol{t}\mid \tilde{\boldsymbol{w}}$}{w,t given w.tilde}}\label{section:bidirectional}
By definition (see equation \ref{eq:t-r}), we have that
\begin{equation*}
	w_h=\tilde{w}_{t_h},
\end{equation*}
hence $\boldsymbol{w}$ differs from $\tilde{\boldsymbol{w}}$ only in relation to the order of its elements, given by $\boldsymbol{t}$. It is therefore only required to sample $\boldsymbol{t}\mid\tilde{\boldsymbol{w}}$ at this step, which implicitly determines $\boldsymbol{w}\mid\tilde{\boldsymbol{w}}$ too.

As $\boldsymbol{w}$ is invariant under size-biased permutations (see \cite{pitman_random_1996} and section \ref{section:sizeBiased} of this article), and as $\tilde{\boldsymbol{w}}$ is a size-biased permutation of it, we have that
\begin{equation*}
	\left(\boldsymbol{w}\mid\tilde{\boldsymbol{w}}=\boldsymbol{\omega}\right) \,{\buildrel d \over =}\, \left(\tilde{\boldsymbol{w}}\mid\boldsymbol{w}=\sigma\left(\boldsymbol{\omega}\right)\right),
\end{equation*} 
for any feasible vector of values $\boldsymbol{\omega}$ and for any permutation $\sigma$ (including the identity permutation). Intuitively, both $\boldsymbol{w}\mid\tilde{\boldsymbol{w}}=\boldsymbol{\omega}$ and $\tilde{\boldsymbol{w}}\mid\boldsymbol{w}=\sigma\left(\boldsymbol{\omega}\right)$ are size-biased random permutations of the same unordered set of weights. Sampling $\boldsymbol{t}\mid\tilde{\boldsymbol{w}}$ therefore involves the same process as sampling $\tilde{\boldsymbol{w}}\mid\boldsymbol{w}$, via size-biasing through equation \ref{eq:sizeBias2}:
\begin{align*}
	&p\left(t_1=\tau\mid\tilde{\boldsymbol{w}}\right)=\tilde{w}_{\tau}, \quad \tau=1,2,\ldots,\\
	&\begin{multlined}p\left(t_j=\tau\mid t_1,\ldots,t_{j-1},\tilde{\boldsymbol{w}}\right)\\=\frac{\tilde{w}_{\tau} \ \mathbbm{1}\left[\tau \neq t_i, \ \textrm{for } 1\leq i \leq {j-1}\right]}{1-\tilde{w}_{t_1}-\ldots-\tilde{w}_{t_{j-1}}}, \quad j>1,\tau=1,2,\ldots\end{multlined}
\end{align*}

\subsubsection{Obtaining \texorpdfstring{$\boldsymbol{r}\mid\boldsymbol{w},\boldsymbol{t},\tilde{\boldsymbol{w}},\boldsymbol{s}$}{r given w,t,w.tilde,s}}
As above, $\boldsymbol{r}^\star\mid\ldots$ can be retrieved from :
\begin{equation}\label{eq:transcoderKey}
	r^\star_j=\min\left\{i:t_i=j\right\}, \quad j=1,2,\ldots
\end{equation}
Then, $\boldsymbol{r}\mid\ldots$ can be derived from $\boldsymbol{s},\boldsymbol{r}^\star$.

\subsection{Testing the transcoding algorithm}
To test the capabilities of the posterior augmentation method, we carry out the following experiment:
\begin{itemize}
	\item we sample a vector $\boldsymbol{r}$ of length 5 from the stick-breaking process with parameter $\alpha=1$, 1 million times;
	\item we re-encode all sample paths of $\boldsymbol{r}$ into $\boldsymbol{s}$, as described in section \ref{section:r2s-1}, and we discard from $\boldsymbol{s}$ and $\boldsymbol{r}$ all sample paths where $\boldsymbol{s}\neq \left(1,1,1,1,2\right)$. By doing so, we are left with $50,090$ data points from a sample from the probability distribution $p\left(\boldsymbol{r}\mid\boldsymbol{s}=\left(1,1,1,1,2\right)\right)$;
	\item we run the posterior augmentation algorithm on $\boldsymbol{s}=\left(1,1,1,1,2\right)$, to infer back $\boldsymbol{r}\mid\boldsymbol{s}=\left(1,1,1,1,2\right)$, and we compare with the frequencies obtained as above. Clearly, the two should match. We compare both marginal and joint frequencies.
\end{itemize}
Our results are in table \ref{table:rFromS-test}. As expected, the probabilities match.

\begin{table*}[t]
	\centering
	\begin{tabular}{lrr}
		&\multicolumn{2}{c}{$p\left(r_1=h\mid\boldsymbol{s}=\left(1,1,1,1,2\right)\right)$}\\
		$h$ & transcoding & empirical\\ \hline
		1  &0.6660 &0.6670\\
		2  &0.2449 &0.2432\\
		3  &0.0677 &0.0682\\
		4  &0.0162 &0.0164\\
		5  &0.0039 &0.0040\\
		6  &0.0009 &0.0010\\
		7  &0.0003 &0.0001\\
		8  &0.0001 &0.0000\\
		$\ldots$&$\ldots$&$\ldots$
	\end{tabular}
	\quad
	\begin{tabular}{lrr}
		&\multicolumn{2}{c}{$p\left(r_5=h\mid\boldsymbol{s}=\left(1,1,1,1,2\right)\right)$}\\
		$h$ & transcoding & empirical\\ \hline
		1  &0.1659 &0.1666\\
		2  &0.3592 &0.3638\\
		3  &0.2281 &0.2286\\
		4  &0.1219 &0.1184\\
		5  &0.0635 &0.0604\\
		6  &0.0304 &0.0306\\
		7  &0.0156 &0.0154\\
		8  &0.0080 &0.0079\\
		$\ldots$&$\ldots$&$\ldots$
	\end{tabular}
	\quad
	\begin{tabular}{lrr}
		&\multicolumn{2}{c}{$p\left(\boldsymbol{r}=\boldsymbol{h}\mid\boldsymbol{s}=\left(1,1,1,1,2\right)\right)$}\\
		$\boldsymbol{h}$ & transcoding & empirical\\ \hline
		$\left(1,1,1,1,2\right)$  &0.3316 &0.3350\\
		$\left(1,1,1,1,3\right)$  &0.1671 &0.1679 \\
		$\left(2,2,2,2,1\right)$  &0.1326 &0.1336 \\
		$\left(1,1,1,1,4\right)$  &0.0838 &0.0823\\
		$\left(2,2,2,2,3\right)$  &0.0561 &0.0560\\
		$\left(1,1,1,1,5\right)$  &0.0426 &0.0401\\
		$\left(2,2,2,2,4\right)$  &0.0280 &0.0265\\
		$\left(3,3,3,3,1\right)$  &0.0266 &0.0268\\
		$\ldots$&$\ldots$&$\ldots$
	\end{tabular}
	\caption{Marginal and joint posterior distribution of $\boldsymbol{r}\mid\boldsymbol{s}=\left(1,1,1,1,2\right)$, obtained via $1$ million simulations from $\boldsymbol{r}$ (``empirical'') and via $100,000$ iterations from the transcoding sampler (``transcoding''), for $\alpha=1$.}
	\label{table:rFromS-test}	
\end{table*}

\section{The transcoding sampler}\label{sec:tranSampler}
We move to the task of developing a sampler for the joint posterior of all of the parameters of interest in the stick-breaking construction of a Dirichlet process mixture: $\boldsymbol{r}$, $\boldsymbol{w}$ and $\boldsymbol{m}$. We write and factor the full joint posterior as:
\begin{align*}
	&p\left(\boldsymbol{r},\boldsymbol{w},\boldsymbol{m},\boldsymbol{s}\mid\boldsymbol{y}\right)\\
	&\quad=p\left(\boldsymbol{s}\mid\boldsymbol{y}\right)\cdot p\left(\boldsymbol{r}\mid\boldsymbol{s},\boldsymbol{y}\right)\cdot p\left(\boldsymbol{w}\mid\boldsymbol{r},\boldsymbol{s},\boldsymbol{y}\right)\cdot p\left(\boldsymbol{m}\mid\boldsymbol{w},\boldsymbol{r},\boldsymbol{s},\boldsymbol{y}\right)\\
	&\quad=p\left(\boldsymbol{s}\mid\boldsymbol{y}\right)\cdot p\left(\boldsymbol{r}\mid\boldsymbol{s},\boldsymbol{y}\right)\cdot p\left(\boldsymbol{w}\mid\boldsymbol{r},\boldsymbol{y}\right)\cdot p\left(\boldsymbol{m}\mid\boldsymbol{r},\boldsymbol{y}\right).
\end{align*}
The first factor can, in principle, be obtained with any sampler, including collapsed and sequential importance samplers (see section \ref{section:postSamplers}). The second factor can be obtained with the transcoding algorithm, as
\begin{equation*}
	p\left(\boldsymbol{r}\mid\boldsymbol{s},\boldsymbol{y}\right)=\frac{p\left(\boldsymbol{y}\mid\boldsymbol{r},\boldsymbol{s}\right)p\left(\boldsymbol{r}\mid\boldsymbol{s}\right)}{p\left(\boldsymbol{y}\mid\boldsymbol{s}\right)}=p\left(\boldsymbol{r}\mid\boldsymbol{s}\right),
\end{equation*}
which the transcoding algorithm is capable of producing. The third factor is also a by-product of the transcoding algorithm.

Hence the transcoding algorithm can be used as a building block to form the \textit{transcoding sampler}, where the full joint conditional is:
\begin{align}\label{eq:factors}
	&p\left(\boldsymbol{r}^\star,\boldsymbol{w},\boldsymbol{m},\boldsymbol{t},\tilde{\boldsymbol{w}},\boldsymbol{s}\mid\boldsymbol{y}\right) \nonumber\\
	&=p\left(\boldsymbol{s}\mid\boldsymbol{y}\right)\cdot p\left(\boldsymbol{r}^\star,\boldsymbol{w},\boldsymbol{t},\tilde{\boldsymbol{w}}\mid\boldsymbol{s}\right)\cdot p\left(\boldsymbol{m}\mid\boldsymbol{r}^\star,\boldsymbol{s},\boldsymbol{y}\right),
\end{align}
where the first factor can be produced with any partition posterior sampler, the second factor can be produced with the transcoding algorithm, and the third factor only requires remapping from $\boldsymbol{\theta}\mid\boldsymbol{s},\boldsymbol{y}$ to $\boldsymbol{m}\mid\boldsymbol{r},\boldsymbol{y}$, or otherwise if absent it can be sampled with standard methods. We henceforth refer to the sampling algorithm for $\boldsymbol{s}\mid\boldsymbol{y}$ as the \textit{core sampler} of the transcoding sampler. Not only can we use as a core sampler one that produces $\boldsymbol{s}\mid\boldsymbol{y}$ directly, but we can also use any other sampler that produces $\boldsymbol{r}\mid\boldsymbol{y}$, which can always be mapped back to $\boldsymbol{s}\mid\boldsymbol{y}$ via the method outlined in section \ref{section:r2s-1}.

The third factor in equation \ref{eq:factors}
is often already available from the core sampler in the first place, in the form of $\boldsymbol{\theta}\mid\boldsymbol{s},\boldsymbol{y}$ (see for example collapsed algorithm 2 and algorithm 8 mentioned in section \ref{section:postSamplers}). For occupied clusters we have
\begin{equation}\label{eq:transcoder-m1}
	m_{r_i}=\theta_i, \quad i=1,\ldots,n,
\end{equation} 
while for unoccupied clusters we have
\begin{equation}\label{eq:transcoder-m2}
	m_j \sim G_0, \quad j\notin \left\{r_1,\ldots,r_n\right\}.
\end{equation}
In case the core sampler does not return the posterior of $\boldsymbol{m}$, we observe that
\begin{equation*}
	p\left(\boldsymbol{m}\mid\boldsymbol{r}^\star,\boldsymbol{s},\boldsymbol{y}\right)=p\left(\boldsymbol{m}\mid\boldsymbol{y},\boldsymbol{r}\right)\propto p\left(\boldsymbol{y}\mid\boldsymbol{m},\boldsymbol{r}\right) p\left(\boldsymbol{m}\right),
\end{equation*}
which can be obtained via standard methods (for example, via Metropolis-Hastings).

To summarise, the transcoding sampler is composed of the following steps:
\begin{enumerate}
	\item use the core sampler to generate posterior samples from $\boldsymbol{s}\mid\boldsymbol{y}$ (this can be any sampler that produces the partition posterior; see for example section \ref{section:postSamplers});
	\item use the transcoding algorithm to sample from $\boldsymbol{r},\boldsymbol{w},\boldsymbol{t},\tilde{\boldsymbol{w}}\mid\boldsymbol{s}$;
	\item if the core sampler also provides the posterior of $\boldsymbol{\theta}$, obtain $\boldsymbol{m}\mid\boldsymbol{\theta},\boldsymbol{r},\ldots$ via equations \ref{eq:transcoder-m1} and \ref{eq:transcoder-m2}, or otherwise sample $\boldsymbol{m}\mid\boldsymbol{y},\boldsymbol{r}$ via other standard approaches (possibly Metropolis-Hastings).
\end{enumerate}

\begin{figure*}
	\centering
	\includegraphics[width=\textwidth]{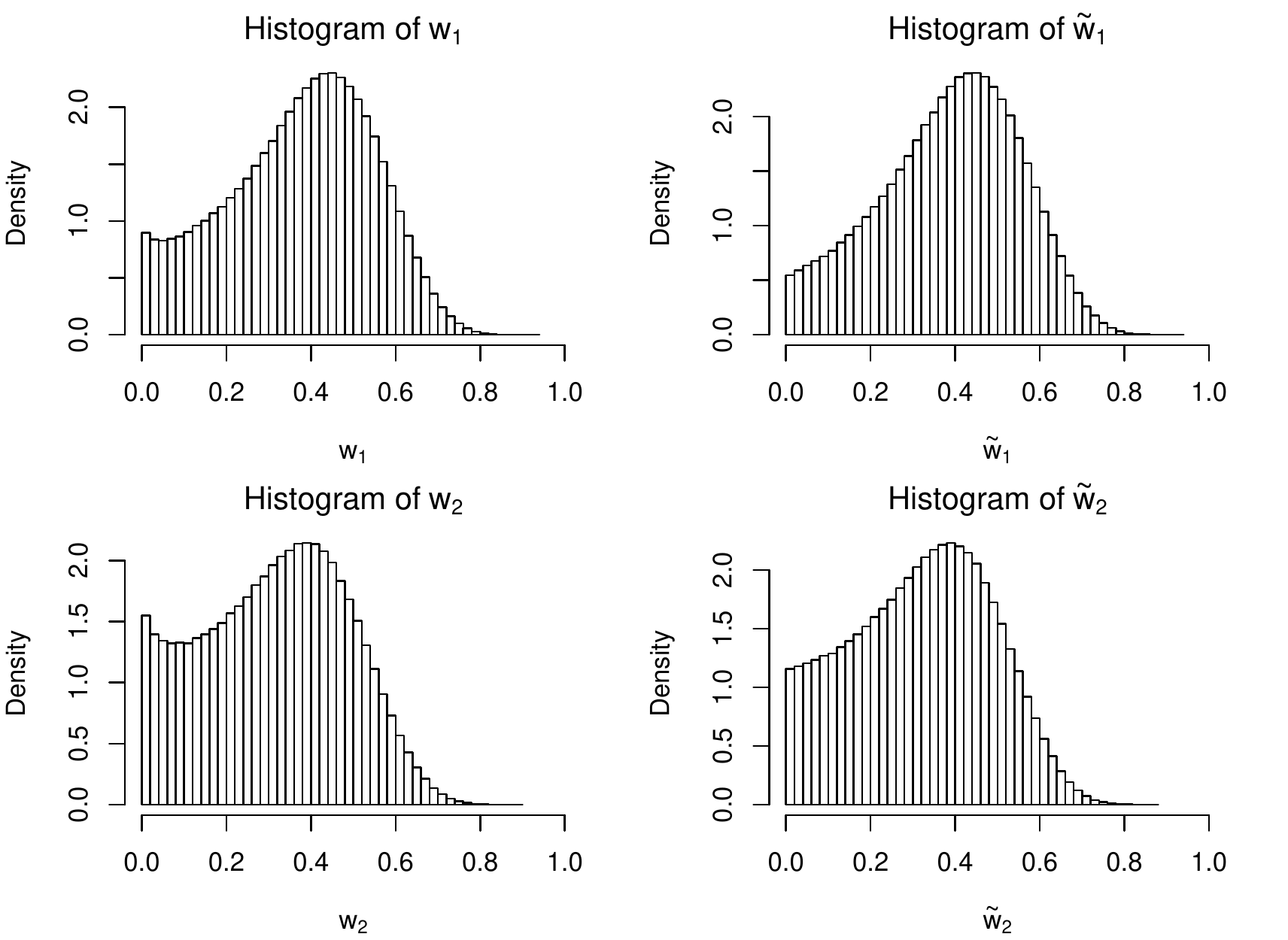}
	\caption{Histogram of the posterior of $w_1,\tilde{w}_1,w_2,\tilde{w}_2$, obtained with the collapsed algorithm 2 and the transcoding sampler, with $\alpha=1$ and 2,000,000 iterations, on the thumb tack data set.}
	\label{figure:wtilde}
\end{figure*}

To stress the difference between $\boldsymbol{w}\mid\boldsymbol{y}$ and $\tilde{\boldsymbol{w}}\mid\boldsymbol{y}$, we include figure \ref{figure:wtilde} as obtained on the testing data set that we describe in section \ref{sec:dataset}: while the a priori probability distribution of $\boldsymbol{w}$ and $\tilde{\boldsymbol{w}}$ is the same, their posterior distributions are not.

\subsection{Performance testing}\label{section:postSamplers}
In this section we carry out performance testing to see how various samplers compare. In what follows, we describe the data set used for testing, the performance measures used for testing, the sampling algorithms subject to testing and the testing outcomes.

\subsubsection{The data set}\label{sec:dataset}
The test data set that we adopt is the thumb tack data of \cite{beckett1994spectral}, in the same order as it appears in \cite{liu1996nonparametric}.  The data set is composed of $320$ observations, pertaining to the roll of a thumb tack; each tack was flipped $9$ times, and ``a one was recorded if the tack landed point up''. We model it via a Dirichlet process mixture with a binomial likelihood, and we set $G_0$ to a $\textrm{Beta}\left(1,1\right)$. For testing purposes, we prefer to fix $\alpha=1$ as opposed to assigning a prior on it, to make it easier to appraise convergence; calculations for $\alpha=0.2$ and $\alpha=5$ (not included here for brevity) broadly lead to the same conclusions, and the methods we outline in this article remain applicable even when $\alpha$ is random. We use native R \citep{rproject} to perform $2,000,000$ iterations of each algorithm; calculations were run on an AMD Ryzen 7 7700X CPU.

\subsubsection{Efficiency measures}\label{section:efficiency}
It is well known that samples from MCMC chains suffer to various degrees from autocorrelation, which makes them less efficient than an equally-sized sample of independent observations. This is well-articulated in \cite{dewitt-morette_monte_1997}, which discusses the concept of \textit{integrated autocorrelation time} of an observable $f$, denoted by $\tau$:
\begin{equation*}
	\tau=\frac{1}{2}\sum_{l=-\infty}^\infty \rho_l=\frac{1}{2}+\sum_{l=1}^\infty \rho_l,
\end{equation*}
where $\rho_l$ is the autocorrelation of $f$ at lag $l$. Measuring $\tau$ is important when comparing MCMC algorithms because $2\tau$ measures how much larger the variance of the estimate of $f$ is than in independent sampling. Like other studies before ours \citep{neal2000, green_modelling_2001, retrospective, walker2011}, we also turn to $\tau$ to gauge the efficiency of the algorithms discussed in this chapter.

In an MCMC chain with $N$ iterations, producing $\hat{\tau}$, the sample estimate of $\tau$, requires the introduction of a cutoff $M\ll N$ in the summation, with ``$M$ large enough so that $\rho_l$ is negligible for $\lvert l \rvert > M$''; this is due to reasons outlined in \cite{dewitt-morette_monte_1997}, and which in essence pertain to the fact that, otherwise, the variance of $\hat{\tau}$ would not converge to $0$ for $N\rightarrow \infty$, which is undesirable behaviour for an estimator. We therefore use the formula
\begin{equation*}
	\hat{\tau}=\frac{1}{2}+\sum_{l=1}^M \hat{\rho}_l,
\end{equation*}
where $M:= \min\left\{l\in\mathbb{N}:l\geq 10 \cdot \hat{\tau}\left(l\right)\right\}$, as per the advice in \cite[page 145]{dewitt-morette_monte_1997}.

The functionals that we turn to, to measure the integrated autocorrelation time, are:
\begin{itemize}
	\item $K_n$, the number of clusters. We choose this metric for comparability, as it has been widely employed in previous studies. Although $K_n$ is not directly altered by a successful label switch, its integrated autocorrelation time should still be informative of the overall convergence speed;
	\item the deviance as defined in \cite{neal2000, green_modelling_2001, retrospective, walker2011}:
	\begin{equation}
		D=-2\sum_{i=1}^n\log\left\{\sum_j \frac{n_j}{n} p\left(y_i\mid\theta_j\right)\right\},	
	\end{equation}
	Previous studies justify its use on the basis that it is seen as ``a global function of all model parameters'' \citep{walker2011};
	\item $\theta_{1}$, which was introduced in \cite{neal2000}. Theoretically, since our testing data set is composed of 320 observations constituted of only 9 unique values, one could monitor $\theta_i$ for nine choices of $i$, corresponding to each unique value of $\boldsymbol{y}$, however for brevity we only report information about $\theta_1$;
	\item $r_1$, which was never considered in previous studies but which in our view is important, since the problem we are trying to solve pertains to \textit{label switching};
	\item $w_1$ and $m_1$ too, for completeness.
\end{itemize}

Two of the samplers that we employ in our tests are sequential importance samplers (SIS) \citep{myOwn,maceachern_sequential_1999}. Integrated autocorrelation time is not the right efficiency measure to compare between sequential importance samplers, which generate i.i.d. observations. To compare between SIS algorithms, we use the variance of their normalised weights, and the Effective Sample Size (ESS), calculated as
\begin{equation*}
	ESS\approx\frac{N}{1+\mathbb{V}\textrm{ar}\left(\hat{\boldsymbol{W}}\right)},
\end{equation*}
where the weight estimates $\hat{W}$ are standardised to sum to 1, as follows:
\begin{equation*}
	\hat{W}_i=\frac{W_i N}{\sum_{i=1}^N W_i},
\end{equation*}
where $N$ is the size of the simulation sample. For more information pertaining to this performance measure, we refer to \cite{liu1996nonparametric,maceachern_sequential_1999}.

\subsubsection{Testing candidates}
In this section we provide a short outline of three broad families of DPM posterior samplers that we compare in our tests, either in conjunction with the transcoding algorithm, or in isolation.

Firstly, we mention the \textit{collapsed} sampler, which owns its name to its use of the Polya urn representation, which integrates $G$ out, therefore \textit{collapsing} the parameter space. This reduction of the parameter space means that the collapsed sampler is less susceptible to slow mixing than stick-breaking samplers; however, it does not allow posterior inference on the random measure $G$ or on other stick-breaking parameters either (as it encodes cluster membership in order of appearance). A comprehensive reference for collapsed samplers is \cite{neal2000}. Neal's algorithm 2 and 8 are generally thought to be the best performing collapsed samplers when the pair $G_0$ and $p\left(y_i\mid\theta_i\right)$ is respectively conjugate or non-conjugate.  Because the test data set that we employ naturally fits with the beta-binomial conjugate pair, we only include in our tables Neal's algorithm 2.

Secondly, we turn our attention to stick-breaking samplers, and to the \textit{slice sampler} in particular. It relies on the DP representation from equation \ref{eq:sethuraman}, and it adopts the stick-breaking encoding. It does provide posterior inference on a number of stick-breaking parameters which the collapsed sampler simply does not provide, such as  $\boldsymbol{w}$ for example. Because of its larger parameter space, it is known that this sampler can suffer from slow mixing of the MCMC chain and as such, some Metropolis jumps have been proposed in literature, to accelerate it. In the tables from this article, we employ the slice sampler \citep{walker2007} as modified in \cite{papaspiliopoulos_note_2008}, accelerated with label-switching Metropolis moves 1 and 2 \citep{retrospective}, and with move 3 \citep{hastie_sampling_2015}.

Thirdly, we mention sequential importance sampling (SIS). In the context of the DPM, it was first explored by \cite{liu1994collapsed}, who introduced a sampler for $\boldsymbol{s},\boldsymbol{\theta}\mid\boldsymbol{y}$. A more efficient variation of it, S2, was introduced shortly thereafter \cite{maceachern_sequential_1999}, which integrates $\boldsymbol{\theta}$ out. Both algorithms encode cluster membership in order of appearance. More recently, we developed SIS algorithm R \citep{myOwn}, which uses stick-breaking encoding instead. The appeal of sequential importance sampling is that it is \textit{i.i.d.}, hence by definition its IAT is the shortest possible; its disadvantage is that its performance depends on the order whereby the observations appear in the sample, with no known optimality criterion to sort the data before the sampler is run. For the purpose of this article, we focus on algorithm S2, which we combine with the transcoding algorithm, and on algorithm R to compare; algorithm R is run standalone rather than in conjunction with the transcoding algorithm because it already uses stick-breaking encoding hence it does not need transcoding.

\subsubsection{Testing outcomes}\label{sec:performance}
Our testing outcomes are reported in \ref{table:conclusions}. The table includes (from top to bottom):
\begin{itemize}
	\item the transcoding sampler with SIS sampler S2 as its core sampler;
	\item SIS algorithm R;
	\item the transcoding sampler with collapsed sampler 2 as its core sampler;
	\item the slice sampler in four variations: without Metropolis moves for acceleration of the chain, and with respectively Metropolis move 1, 2 or 3.
\end{itemize}
It should be highlighted that the transcoding sampler is guaranteed to inherit its IAT from its core sampler, by construction; for example, although we do not include the IAT of stand-alone collapsed algorithm 2 in \ref{table:conclusions}, it is safe to assume that its IAT is exactly the same as that of the transcoding sampler that uses collapsed algorithm 2 as its core sampler.

As expected, where the core sampler is i.i.d., the transcoding sampler returns IAT values at the minimum end of the spectrum (i.e. $\approx 0.50$) -- it produces fully i.i.d. samples. Where the core sampler is instead a collapsed algorithm, the IAT of the parameters that are naturally produced by the core sampler is by construction exactly that of the core sampler, and the IAT of the stick-breaking parameters which result from the augmentation of the core sampler displays materially shorter IAT than that of the slice sampler, no matter which Metropolis label switching move the latter adopts.

\begin{table*}[t]
	\centering
	\makebox[\textwidth][c]{
		\begin{tabular}{lrrrrrrrr}
			Algorithm & ESS & $\textrm{IAT}_K$ & $\textrm{IAT}_{w_1}$ & $\textrm{IAT}_{r_1}$ & $\textrm{IAT}_{w_{r_1}}$ & $\textrm{IAT}_{m_1}$&$\textrm{IAT}_{\theta_1}$ & $IAT_D$\\ \hline
			SIS S2+transcoding& 144,211& 0.50 & 0.50 & 0.50 & 0.50& 0.50&0.50&0.50 \\
			SIS R& 131,933& 0.50 & 0.50 & 0.50 & 0.50&	0.50& 0.50& 0.50\\
			collapsed2+transc.&N.A.&11.86&5.97&2.49& 7.73 & 0.50&0.55&	2.15\\
			slice, no moves &N.A.& 75.16 & 126.00 & 43.70 & 36.00 & 388.12 & 0.87 & 6.43\\
			slice, move 1 &N.A.& 72.16 & 121.15 & 32.15 & 34.34 & 38.69 & 0.87 & 6.65\\
			slice, move 2 &N.A.& 65.39 & 73.59 & 32.96 & 32.44 & 226.31 & 0.86 & 6.39\\
			slice, move 3 &N.A.& 57.37 & 34.38 & 13.64 & 29.63 & 6.15 & 0.86 & 6.24
		\end{tabular}
	}
	\caption{Comparison of the transcoding sampler with sequential importance sampling algorithm S2 (``SIS S2+transcoding'') and collapsed algorithm 2 (``collapsed2+transc.'') as its core samplers, and the slice sampler and sequential importance sampling algorithm R, over $2,000,000$ iterations, on the thumb tack data set, with $\alpha=1$.}
	\label{table:conclusions}	
\end{table*}

\section{Relationship with other work}\label{sec:compare}
Ultimately, the three main building blocks of the transcoding sampler are:
\begin{enumerate}
	\item \label{bb1} size-biasing;
	\item \label{bb2} equation \ref{eq:pitman0};
	\item \label{bb3} the derivation of $\boldsymbol{w},\boldsymbol{t}\mid\tilde{\boldsymbol{w}}$, and equation \ref{eq:transcoderKey}.
\end{enumerate}

We have identified two other approaches which revolve around building blocks \ref{bb1} and \ref{bb2}, yet they reach different conclusions from those outlined in this article, as they do not use building block \ref{bb3}.

One is \cite{fall:hal-00740770}, where an algorithm is developed which, in essence, samples from  
\begin{equation}\label{eq:fall2}
	p\left(s_i\mid\tilde{\boldsymbol{w}},\boldsymbol{\theta}^\star\ldots\right)\propto \tilde{w}_{s_i} p\left(y_i\mid\theta^\star_{s_i}\right),\quad i=1,\ldots,n,
\end{equation}
then  it samples $\tilde{\boldsymbol{w}}\mid\boldsymbol{s},\ldots$ via equation \ref{eq:fall}, and then it samples their locations, in a Gibbs scheme, where cluster membership indicators are re-encoded at each iteration so that they are in order of appearance. The space that this algorithm operates in is somewhere in between that of collapsed algorithm 2, and that of the slice sampler: it is wider than the former, because it also includes $\tilde{\boldsymbol{w}}$, while it is narrower than the latter, as its cluster membership encoding is in order of appearance. Its IAT times published in \cite{fall:hal-00740770} reflect the same, positioning the algorithm between collapsed algorithm 2 and the slice sampler in terms of performance. However, the posterior that this algorithm produces does not include either $\boldsymbol{r}$ or $\boldsymbol{w}$, which instead the slice sampler returns -- as such, contrary to the transcoding sampler, it is not a replacement for the slice sampler.

The other is \cite{doi:10.1080/10618600.2023.2177298}\footnote{Also related to \cite{leyva}.}, which essentially\footnote{The algorithm also has other features and uses, including for example its applicability to finite mixture models.} also operates as indicated earlier, with the technical difference that instead of sampling $s_i\mid\ldots$ from equation \ref{eq:fall2} and re-encoding the labels in order of appearance at each iteration, it adds constraints to the sampling formula so that the resulting cluster membership indicators generated by the sampler are always admissible to begin with, by construction, hence no need to re-encode them. This algorithm operates in the same space as \cite{fall:hal-00740770}, and as above its posterior does not include $\boldsymbol{r}$ or $\boldsymbol{w}$ either, and is no substitute for the slice sampler.

\section{Conclusions}\label{sec:conclusions}
In this article, we have discussed cluster membership encoding according to the order of appearance of the clusters in the data (as it happens in the vector that we denote by $\boldsymbol{s}$, which results from the Polya urn construction), and according to the order of their originating sticks in the stick-breaking construction process (as it happens in the vector that we denote by $\boldsymbol{r}$, which results from the stick-breaking construction); we have determined that the latter carries more information than the former. We have worked out a simple, deterministic approach to derive $\boldsymbol{s}\mid\boldsymbol{r}$, and the \textit{transcoding algorithm} to infer $\boldsymbol{r}\mid\boldsymbol{s}$.

We have also derived the \textit{transcoding sampler}, which fully integrates with any other sampler capable of returning the exchangeable posterior partition of the data (i.e. its \textit{core sampler}), to infer back all stick-breaking parameters of possible interest including $\boldsymbol{r},\boldsymbol{w},\tilde{\boldsymbol{w}}$ and others. Since the transcoding sampler leverages the transcoding algorithm, which is i.i.d., it inherits the integrated autocorrelation times of the core sampler that it relies on. The transcoding sampler thus makes it possible to make full posterior inferences of stick-breaking parameters while attaining minimal autocorrelation times (when using the sequential importance sampler at its core), or while attaining the same autocorrelation times as those from collapsed samplers (when using collapsed samplers at its core), which are known to be much shorter than those attained by the slice sampler and other stick-breaking samplers (even in the presence of Metropolis label-switching moves, to accelerate the slice sampler).

\appendix

\section{A new a prior sampler for $\boldsymbol{s}$}\label{sec:bridge}
For better understanding of the connection between the two encodings, in this appendix we exemplify the relationship between $\boldsymbol{w}\mid\tilde{\boldsymbol{w}}=\boldsymbol{\omega}$ and $\tilde{\boldsymbol{w}}\mid\boldsymbol{w}=\sigma\left(\boldsymbol{\omega}\right)$ that we made use of in section \ref{section:bidirectional}, and we show that each is a size-biased random permutation of the other, in both directions.  To do so, we introduce a new \textit{a priori} sampler for $\boldsymbol{s}$ which does not rely on the classic Polya urn construction of equation \ref{eq:polya10}. The process is as follows:
\begin{enumerate}
	\item set $s_1=1$ and sample $\tilde{w}_1\sim \textrm{Beta}\left(1,\alpha\right)$ as per equation \ref{eq:sethuraman} (this is possible because $\tilde{\boldsymbol{w}}$ is a size-biased permutation of $\boldsymbol{w}$ and the latter is invariant to size-biased permutations; see \cite{pitman_random_1996} and \cite{hoppe_sampling_1987}). Then, sample $s_2, s_3, \ldots$ from:
	\begin{multline}\label{eq:complex}
		p\left(s_i\mid s_1,\ldots,s_{i-1},\tilde{\boldsymbol{w}}\right)\\=\left(\sum_{j=1}^{k_{i-1}}\tilde{w}_j\delta_j\left(s_i\right)\right) + \left(1-\sum_{l\leq k_{i-1}}\tilde{w}_l\right)\delta_{k_{i-1}+1}\left(s_i\right),
	\end{multline}
	where $k_{i-1}$ is the number of clusters in $s_1,\ldots,s_{i-1}$. 	We derived equation \ref{eq:complex} from equation 3.4 from \cite{pitman_combinatorial_2002}:
	\begin{equation}\label{eq:pitman000}
		p\left(\tilde{n}_1,\ldots,\tilde{n}_k \mid\tilde{\boldsymbol{w}}\right)=\left(\prod_{j=1}^k \tilde{w}_j^{\tilde{n}_j-1}\right) \left(\prod_{j=1}^{k-1} 1-\sum_{l\leq j} \tilde{w}_l\right).
	\end{equation}
	Intuitively, given $s_1,\ldots,s_{i-1}$ and $\tilde{w}_1,\ldots,\tilde{w}_{k_{i-1}}$, the probability of $s_i$ opening up a new cluster is $1-\sum_{l\leq k_{i-1}}\tilde{w}_l$ (see the second factor in equation \ref{eq:pitman000}); conversely, the probability of $s_i$ repeating some already observed cluster $j$ is $\tilde{w}_j$ (see the first factor in the same equation).  Whenever a new cluster is opened up, draw its corresponding weight as per equation \ref{eq:sethuraman}, on the basis of the same justification as before.
	\item draw a size-biased permutation of $\tilde{\boldsymbol{w}}$, to obtain $\boldsymbol{w}$ and implicitly $\boldsymbol{r}$ too, by using a variation of equation \ref{eq:sizeBias2} where, in the formula, $\boldsymbol{w}$ is replaced by $\tilde{\boldsymbol{w}}$ and vice-versa.
\end{enumerate}

This sampler is noteworthy because of two reasons:
\begin{itemize}
	\item it generates data points from the Polya urn process in a way that is different from how it is usually presented. This method allows the weights to be exact, as opposed to Polya urn samplers based on equation \ref{eq:polya10} where their estimation is based on the size of the generated clusters, and is only asymptotically correct;
	\item it exemplifies that $\boldsymbol{w}$ and $\tilde{\boldsymbol{w}}$ are size-biased permutations of each other.  While $\tilde{\boldsymbol{w}}$ naturally arises as a size-biased random permutation of $\boldsymbol{w}$ due to its connection to sampling without replacement, $\boldsymbol{w}$ can also be thought as a size-biased random permutation of $\tilde{\boldsymbol{w}}$.
\end{itemize}

\section{Exchangeability, EPPF and Ewens' sampling formula}
We include some considerations about the concept of exchangeability as it applies to random vectors and to random partitions, also useful to support some conclusions in appendix \ref{appendix:AR}.

\subsection{Exchangeability}
One of the main properties generally associated with the Polya urn construction of the Dirichlet process is that it produces ``exchangeable sequences'' and, while this is true under certain conditions, we would like to clarify the impact that different types of encoding have on this key property.

The finite Polya urn scheme, with a fixed number colours, is known to generate a sequence of random cluster membership indicators that is exchangeable (in the sense of de Finetti), and which is conditionally independent on a Dirichlet probability measure \citep{hill1987exchangeable}, when it is coded according to a finite set with elements corresponding to each colour. Similarly, the infinite-dimensional Polya urn is also known to produce sequences of random cluster membership indicators which are exchangeable, and which are conditionally independent on a Dirichlet process measure, as long as each of their labels uniquely identifies a specific colour.  However, we observe that encoding a Polya urn sequence with the \textit{order of appearance} encoding breaks its de Finetti exchangeability, due to the order constraints that the encoding imposes. For example, assume a Polya urn sequence with parameter $\alpha$: the random vector $\left(s_1,s_2,\ldots\right)$ that it induces is not exchangeable as $s_1=1$ almost surely, while $p\left(s_2=1\right)<1$, hence $p\left(s_1,s_2\right)\neq p\left(s_2,s_1\right)$. It can be seen that the conditional $\boldsymbol{s}\mid\boldsymbol{r}$ is not exchangeable either.

Conversely, stick-breaking encoding does lead to a sequence $\boldsymbol{r}$ that is de Finetti exchangeable, as $r_i,r_j$ are conditionally independent on $\boldsymbol{w}$. However, its conditional $\boldsymbol{r}\mid\boldsymbol{s}$ is only de Finetti partially exchangeable -- it is not fully exchangeable.

What is preserved under both encoding methods is the exchangeability of the random partition induced by the Dirichlet process: both encoding methods lead to a random exchangeable partition.  Recall that a random partition is exchangeable if and only if its probability distribution is symmetric for every permutation $\sigma$ of $\mathbb{N}_k$:
\begin{equation*}
	p\left(n_1,\ldots,n_k\right)=p\left(n_{\sigma_{1}},\ldots,n_{\sigma_k}\right).
\end{equation*}

We summarise these properties in table \ref{table:ex}.

\begin{table}[ht]
	\centering
	\begin{tabular}{lcccc}
		property & $\boldsymbol{s}$ & $\boldsymbol{s}\mid\boldsymbol{r}$ & $\boldsymbol{r}$ & $\boldsymbol{r}\mid \boldsymbol{s}$ \\ \hline
		exchangeability	& no	& no & yes & no\\
		partial exchangeability	& no & no & yes & yes\\
		exchangeability of its partition & yes & yes & yes & yes\\
	\end{tabular}
	\caption{Exchangeability of the cluster membership indicator vector when encoded in order of appearance ($\boldsymbol{s}$) and in stick-breaking order ($\boldsymbol{r}$), and of its posterior.}
	\label{table:ex}	
\end{table}

\subsection{EPPF and Ewens' sampling formula}\label{sec:EPPF4}
An exchangeable random partition $\Pi_n$ can be described by the following function, which is called \textit{exchangeable partition probability function} (EPPF):
\begin{equation}\label{eq:ewens1}
	p\left(n_1,\ldots,n_k\right)=\alpha^k \frac{\Gamma\left(\alpha\right)}{\Gamma\left(\alpha+n\right)}\prod_{j=1}^k \Gamma\left(n_j\right).
\end{equation}
The EPPF is a symmetric function which returns the probability of one particular (unordered) set partition, and which only depends on the unordered block sizes; partitions with the same block sizes have the same probability. 

Equation \ref{eq:ewens1} is also related to the following, which is called the \textit{Ewens sampling formula}. Define $M_j:=\#\left\{i:n_i=j,\ i=1,\ldots,k\right\},\ j=1,\ldots,n$. Then  
\begin{equation}\label{eq:ewens2}
	p\left(M_1,\ldots,M_n\right)=n! \frac{\Gamma\left(\alpha\right)}{\Gamma\left(\alpha+n\right)}\prod_{i=1}^n 
	\frac{\alpha^{M_i}}{i^{M_i} {M_i}!},
\end{equation}
as for every configuration in Equation \ref{eq:ewens2} there are 
\begin{equation*}
	\frac{n!}{\prod_{i=1}^n i!^{M_i}M_i!}
\end{equation*}
configurations in \ref{eq:ewens1} (see  \cite{crane_ubiquitous_2016} for a comprehensive summary of the uses of the Ewens sampling formula). 

Finally, the probability of the composition $\left(n_1,\ldots,n_k\right)$ in order of appearance is \citep[equation 2.6]{pitman_combinatorial_2002}:
\begin{align}
	p_{\textrm{OOA}}\left(n_1,\ldots,n_k\right)&=\frac{n!}{n_k\left(n_k+n_{k-1}\right)\cdots\left(n_k+\cdots+n_1\right)}\ \frac{p\left(n_1,\ldots,n_k\right)}{\prod_{i=1}^k \left(n_i-1\right)!}\nonumber \\
	&=\frac{n!}{n_k\left(n_k+n_{k-1}\right)\cdots\left(n_k+\cdots+n_1\right)}\ \alpha^k \frac{\Gamma\left(\alpha\right)}{\Gamma\left(\alpha+n\right)}.\label{eq:pOOA}
\end{align}

\section{Accept-reject transcoding methods}\label{appendix:AR}
The simplest way to obtain stick-breaking encoding from cluster membership indicators encoded in order of appearance is via the accept/reject algorithm. We present here variations of it, in increasing order of efficiency; however, even in its most efficient form the algorithm is still wasteful, hence the value of the \textit{transcoding algorithm} from section \ref{sec:transAlgo} which instead operates via augmentation of the target space.

\subsection{Accept-reject method 1}\label{section:AR-algo1}
To infer $\boldsymbol{r}$ from $\boldsymbol{s}$, we write
\begin{equation}\label{eq:r-indicator}
	p\left(\boldsymbol{r}\mid\boldsymbol{s}\right)\propto p\left(\boldsymbol{s}\mid\boldsymbol{r}\right) p\left(\boldsymbol{r}\right),
\end{equation}
and we observe that:
\begin{itemize}
	\item $p\left(\boldsymbol{r}\right)$ is easy to sample from, by progressively sampling from $p\left(r_1\right), p\left(r_2\mid r_1\right), \ldots$ according to:
	\begin{equation*}
		p\left(r_1=h\right)=\mathbb{E}\left[w_h\right]=\frac{\alpha^{h-1}}{\left(\alpha+1\right)^{h}},
	\end{equation*}
	and, for $i=2,3,\ldots$, according to:
	\begin{align*}
		p\left(r_i=h\mid r_1,\ldots,r_{i-1}\right)&=\mathbb{E}\left[w_h\mid r_1,\ldots,r_{i-1}\right]\\
		&=\mathbb{E}\left[v_h\mid r_1,\ldots,r_{i-1}\right] \prod_{l<h} \mathbb{E}\left[1-v_l\mid r_1,\ldots,r_{i-1}\right],
	\end{align*}
	where the conditional expectations can be obtained from
	\begin{equation}\label{eq:blockedV}
		v_h\mid\ldots\sim \textrm{Beta}\left(1+n_h,\alpha+\sum_{l=h+1}^H n_l\right),
	\end{equation}
	where $n_h$ is the number of observations in cluster $h$, and which in turn is due to the conjugacy of the generalised Dirichlet distribution \citep{connor1969concepts};
	\item the first term on the right hand side of equation \ref{eq:r-indicator} acts as an indicator function:
	\begin{equation*}
		p\left(\boldsymbol{s}\mid\boldsymbol{r}\right) =\mathbbm{1}_{g\left(\boldsymbol{r}\right)}\left(\boldsymbol{s}\right).
	\end{equation*}
\end{itemize}
Therefore, we can draw a proposal $\hat{\boldsymbol{r}}$ from $p\left(\boldsymbol{r}\right)$, and accept it if $g\left(\hat{\boldsymbol{r}}\right)=\boldsymbol{s}$, or repeat the attempt if otherwise. The resulting acceptance rate equals equation \ref{eq:ewens1} (Ewens' distribution), which returns the probability of one specific configuration of $\boldsymbol{s}$ where the cluster sizes are $n_1,\ldots,n_k$:
\begin{equation*}
	\alpha^k \frac{\Gamma\left(\alpha\right)}{\Gamma\left(\alpha+n\right)}\prod_{j=1}^k \Gamma\left(n_j\right).
\end{equation*}
However, the probability above approaches zero very quickly as $n$ increases, to the point of quickly becoming unusable; for example, for a sequence with $n=30,\alpha=1$ with $3$ clusters of size $\left(22,7,1\right)$, its theoretical acceptance rate as derived via equation \ref{eq:ewens1} is approximately $\num{1.4e-10}$, hence the algorithm is quite wasteful.

\begin{table}[ht]
	\centering
	\begin{tabular}{llll}
		Row $$ & $\boldsymbol{s}$ & $p\left(\boldsymbol{s}\right)$ & acceptable $\boldsymbol{r}$ pattern\\ \hline
		$1$ & $\left(1,1,1\right)$ & $0.3333$ & $\left(a,a,a\right)$\\
		$2$ & $\left(1,1,2\right)$ & $0.1666$ & $\left(a,a,b\right)$\\
		$3$ & $\left(1,2,1\right)$ & $0.1666$ & $\left(a,b,a\right)$\\
		$4$ & $\left(1,2,2\right)$ & $0.1666$ & $\left(a,b,b\right)$\\
		$5$ & $\left(1,2,3\right)$ & $0.1666$ & $\left(a,b,c\right)$
	\end{tabular}
	\caption{Accept-reject algorithm $1$, acceptable configurations of $\boldsymbol{r}$ for all possible outcomes of $\boldsymbol{s}$, in a Dirichlet process where $n=3,\alpha=1$.}
	\label{table:permS}	
\end{table}

\subsection{Accept-reject method 2}\label{section:AR-algo2}
Consider a permutation $\sigma$ of $\mathbb{N}$, and observe that:
\begin{itemize}
	\item while $\boldsymbol{s}$ is not de Finetti exchangeable (because of its order constraint), we still have that $p\left(s_1,\ldots,s_n\right)=p\left(s_{\sigma_1},\ldots,s_{\sigma_n}\right)$ as long as $\sigma$ does not alter the order of appearance of the clusters;
	\item $p\left(r_1,\ldots,r_n\right)=p\left(r_{\sigma_1},\ldots,r_{\sigma_n}\right)$, for any $\sigma$, because the random vector $\boldsymbol{r}$ is de Finetti exchangeable.
\end{itemize}

It is therefore legitimate to relax the acceptance criterion from section \ref{section:AR-algo1} to accept all cases where the ordered\footnote{With $\boldsymbol{r}$ encoded in stick-breaking order, and with the sizes of the clusters ordered so that the size of the first cluster to appear is positioned first, the size of the second cluster to appear is positioned second, etc.} vector of the cluster sizes of $\hat{\boldsymbol{r}}$ matches the ordered cluster sizes of $\boldsymbol{s}$; doing so increases the acceptance rate to
\begin{equation*}
	\alpha^k\frac{\Gamma\left(\alpha\right)}{\Gamma\left(\alpha+n\right)}\frac{n!}{n_k\left(n_k+n_{k-1}\right)\cdots\left(n_k+\ldots+n_1\right)},
\end{equation*}
as per equation \ref{eq:pOOA}.

We exemplify this with the aid of tables \ref{table:permS} and \ref{table:permS2}, which shows the set of all possible outcomes of $\boldsymbol{s}$ when $n=3$. When, for example, $\boldsymbol{s}=\left(1,1,2\right)$, the algorithm from section \ref{section:AR-algo1} aims to capture all outcomes of $\boldsymbol{r}$ whose configuration is compatible with $\boldsymbol{s}=\left(1,1,2\right)$; there is only one such case in Table \ref{table:permS} (see row $2$). The algorithm from this section, instead, moves from the observation that row $3$ from Table \ref{table:permS} is entirely equivalent to row $2$ of the same, once both $\boldsymbol{s}$ and $\boldsymbol{r}$ transformed via $\sigma_1=1,\sigma_2=3,\sigma_3=2$.  In this specific example, doing so doubles the acceptance ratio; in the example from section \ref{section:AR-algo1} ($n=30,\alpha=1$; $3$ clusters of size $22,7,1$), it improves the acceptance rate from $\num{1.4e-10}$ to $0.00416$. Table \ref{table:permS2} summarises the configurations of $\boldsymbol{r}$ that can be accepted, once appropriately re-arranged.

\begin{table}[ht]
	\centering
	\begin{tabular}{llll}
		Row $$ & $\boldsymbol{s}$ & $p\left(\boldsymbol{s}\right)$ & acceptable $\boldsymbol{r}$ pattern\\ \hline
		$1$ & $\left(1,1,1\right)$ & $0.3333$ & $\left(a,a,a\right)$\\
		$2$ & $\left(1,1,2\right)$ & $0.1666$ & $\left(a,a,b\right),\left(a,b,a\right)$\\
		$3$ & $\left(1,2,1\right)$ & $0.1666$ & $\left(a,b,a\right),\left(a,a,b\right)$\\
		$4$ & $\left(1,2,2\right)$ & $0.1666$ & $\left(a,b,b\right)$\\
		$5$ & $\left(1,2,3\right)$ & $0.1666$ & $\left(a,b,c\right)$
	\end{tabular}
	\caption{Accept-reject algorithm 2, acceptable configurations of $\boldsymbol{r}$ for all possible outcomes of $\boldsymbol{s}$, in a Dirichlet process where $n=3,\alpha=1$.}
	\label{table:permS2}	
\end{table}

\subsection{Accept-reject method 3}\label{section:AR-algo3b}
The approach from section \ref{section:AR-algo2} can be pushed even further by observing that  $p\left(s_1,\ldots,s_n\right)=p\left(\tau\left(s_{\sigma_1}\right),\ldots,\tau\left(s_{\sigma_n}\right)\right)$, for any permutations $\sigma,\tau$ which do not alter the order of appearance of the clusters in $\left(\tau\left(s_{\sigma_1}\right),\ldots,\tau\left(s_{\sigma_n}\right)\right)$, as $\boldsymbol{s}$ is invariant to permutations of its labels (subject to the aforementioned condition). This is exemplified in Table \ref{table:permS3}.

The algorithm works by generating a proposal $\hat{\boldsymbol{r}}$, sorting its cluster sizes in increasing (or decreasing) order and comparing them with the sorted cluster sizes of $\boldsymbol{s}$; if they match, $\hat{\boldsymbol{r}}$ is accepted and the positions of its elements are permuted to match the pattern of $\boldsymbol{s}$.

The acceptance rate of this algorithm is (see section \ref{sec:EPPF4}):
\begin{equation}\label{eq:factor}
	\frac{n!}{\prod_{i=1}^n M_i!\left(i!\right)^{M_i}} \cdot \alpha^k \frac{\Gamma\left(\alpha\right)}{\Gamma\left(\alpha+n\right)}\prod_{j=1}^k \Gamma\left(n_j\right).
\end{equation}

This is a moderate improvement to the algorithm from section \ref{section:AR-algo2}; for example, in the same case as above ($n=30,\alpha=1$; $3$ clusters of size $22,7,1$), its acceptance rate is $0.0065$, up from $0.00416$. However, as is to be expected, even this algorithm becomes more wasteful as $n$ increases, and in fact, for a sample size of $n=320$ (such as for example the thumbtack dataset used in \cite{liu1996nonparametric}), the acceptance rate is $1$ in about $70$ million, for cluster sizes $\left(226,75,13,3,2,1\right)$, which motivates the need to devise a better algorithm.

\begin{table}[ht]
	\centering
	\begin{tabular}{llll}
		Row $$ & $\boldsymbol{s}$ & $p\left(\boldsymbol{s}\right)$ & acceptable $\boldsymbol{r}$ pattern\\ \hline
		$1$ & $\left(1,1,1\right)$ & $0.3333$ & $\left(a,a,a\right)$\\
		$2$ & $\left(1,1,2\right)$ & $0.1666$ & $\left(a,a,b\right),\left(a,b,a\right),\left(a,b,b\right)$\\
		$3$ & $\left(1,2,1\right)$ & $0.1666$ & $\left(a,b,a\right),\left(a,a,b\right),\left(a,b,b\right)$\\
		$4$ & $\left(1,2,2\right)$ & $0.1666$ & $\left(a,b,b\right),\left(a,a,b\right),\left(a,b,a\right)$\\
		$5$ & $\left(1,2,3\right)$ & $0.1666$ & $\left(a,b,c\right)$
	\end{tabular}
	\caption{Accept-reject algorithm 3, acceptable configurations of $\boldsymbol{r}$ for all possible outcomes of $\boldsymbol{s}$, in a Dirichlet process where $n=3$.}
	\label{table:permS3}	
\end{table}

\bibliography{bibliography}

\end{document}